\documentclass[12pt]{iopart}

\usepackage{graphicx}
\usepackage{amsthm}
\usepackage{amsfonts}
\usepackage{algorithm}
\usepackage{algorithmic}

\usepackage[colorlinks,
            linkcolor=red,
            anchorcolor=red,
            citecolor=red,
            ]{hyperref}

\newtheorem{proposition}{Proposition}
\newtheorem{theorem}{Theorem}
\newtheorem{corollary}{Corollary}
\newtheorem{remark}{Remark}

\def\bt{\begin{theorem}}
\def\et{\end{theorem}}
\def\bq{\begin{question}}
\def\eq{\end{question}}
\def\bp{\begin{proposition}}
\def\ep{\end{proposition}}
\def\bc{\begin{corollary}}
\def\ec{\end{corollary}}
\def\bo{\begin{proof}}
\def\eo{\end{proof}}
\def\bx{\begin{example}}
\def\ex{\end{example}}
\def\br{\begin{remark}}
\def\er{\end{remark}}
\def\bl{\begin{lemma}}
\def\el{\end{lemma}}

\def\bn{\begin{definition}}
\def\en{\end{definition}}
\def\ba{\begin{array}}
\def\ea{\end{array}}
\def\be{\begin{equation}}
\def\ee{\end{equation}}

\def\i{{\bf i}}

\def\r{{\bf r}}

\def\u{{\bf u}}
\def\v{{\bf v}}
\def\w{{\bf w}}
\def\x{{\bf x}}
\def\y{{\bf y}}
\def\z{{\bf z}}

\def\0{{\bf 0}}
\def\1{{\bf 1}}
\def\2{{\bf 2}}
\def\3{{\bf 3}}
\def\4{{\bf 4}}
\def\5{{\bf 5}}
\def\6{{\bf 6}}
\def\7{{\bf 7}}
\def\8{{\bf 8}}
\def\9{{\bf 9}}

\def\ds{\displaystyle}

\begin{document}

\title[A Quantum Model for Multilayer Perceptron]{A Quantum Model for Multilayer Perceptron}

\author{Changpeng Shao}

\address{Academy of Mathematics and Systems Science, Chinese Academy of Sciences, Beijing 100190, China}
\ead{cpshao@amss.ac.cn}

\begin{abstract}
Multilayer perceptron is the most common used class of feed-forward artificial neural network. It contains many applications
in diverse fields such as speech recognition, image recognition, and machine translation software.
To cater for the fast development of quantum machine learning, in this paper, we propose a new model to study multilayer perceptron in quantum computer.
This contains the tasks to prepare the quantum state of the output signal in each layer and to establish the quantum version of learning algorithm about the weights in each layer.
We will show that the corresponding quantum versions can achieve at least quadratic speedup or even exponential speedup over the classical algorithms.
This provide us an efficient method to study multilayer perceptron and its applications in machine learning in quantum computer.
Finally, as an inspiration, an exponential fast learning algorithm (based on Hebb's learning rule) of Hopfield network will be proposed.
\end{abstract}


\section{Introduction}
\setcounter{equation}{0}

Inspired by biological neural networks, artificial neural networks (ANNs) are massively parallel computing models consisting of an extremely large number of simple processors with many interconnections. ANNs are highly successful methods to study machine learning. Researchers from different scientific disciplines are designing different ANN models to solve various problems in pattern recognition, clustering, approximation, prediction, optimization, control and so on \cite{bishop,fukunaga,hornik,hui,jain,schiffman,suykens}.
Hence ANNs are of great interest for quantum adaptation. Quantum neural networks (QNNs) are models combining the powerful features of quantum computing (such as superposition and entanglement) and ANNs (such as parallel computing). However, the nonlinear dynamics of ANNs are very different from unitary operations (which are linear) of quantum computing. To find a meaningful QNN model that integrates both fields is a highly nontrivial task \cite{schuld-qnn}. Many attempts of QNN models are obtained
\cite{altaisky,andrecut,behrman,kak,ricks,schuld-qnn,schuld-qml,silva,siomau,wan}.

In \cite{altaisky}, Altaisky introduced a quantum perceptron which is modelled by the following quantum updating function $|y(t)\rangle=\hat{F}\sum_i \hat{w}_{iy}(t) |x_i\rangle$, where $\hat{F}$ is an arbitrary unitary operator and $\hat{w}_{iy}(t)$ is an operator representing the weights. The training rule is $\hat{w}_{iy}(t+1)=\hat{w}_{iy}(t) +\eta (|d\rangle-|y(t)\rangle) \langle x_i|$, where $|d\rangle$ in the quantum state of the target vector,  $|y(t)\rangle$ is the output state of the quantum perceptron and $\eta$ is some pramteter. The author notes that this training rule is by no means unitary regarding the components of the weight matrix.
So it would fail to preserve the unitary property and so the total probability of the system.

A large class of ANNs uses binary McCulloch-Pitts neurons \cite{McCulloch-Pitts} in which the neural cells are assumed to be active or resting. The value of inputs and outputs are $\{-1,1\}$. So there is a natural connection between the the inputs $\{-1,1\}$ and qubits $|0\rangle,|1\rangle$. All possible inputs $x_1,\ldots,x_n\in \{-1,1\}$ naturally corresponds to a quantum state $\sum_{x_1,\ldots,x_n\in\{-1,1\}} \alpha_{x_1,\ldots,x_n} |x_1,\ldots,x_n\rangle$, where $\alpha_{x_1,\ldots,x_n}$ refers to the amplitude of $|x_1,\ldots,x_n\rangle$. In \cite{schuld-qml}, Schuld et al. considered the simulation of a perceptron in quantum computer in this model. They put the signals $x_1,\ldots,x_n\in\{-1,1\}$ into qubits $|x_1,\ldots,x_n\rangle$, so that the output of a perceptron under the threshold function can be computed to some precision by applying quantum phase estimation on a unitary operator, which is generated by the weights. This model uses $n$ qubits, so the complexity of the method given in \cite{schuld-qml} is linear at $n$. For the introduction of other QNN models, we refer to \cite{schuld-qnn}.

In this paper, we give a try to study multilayer perceptron (MLP) in quantum computer from a different idea, that is setting $x_i$ as the amplitude of qubits to generate the quantum state $|x\rangle=\frac{1}{\|\x\|}\sum x_i|i\rangle$, where $\x=(x_1,\ldots,x_n)$ is the vector of input signals.
This idea has been applied in \cite{rebentrost} to study quantum Hopfield neural network.
One apparent advantage is that it only uses $O(\log n)$ qubits.
So it may have a better performance than the model used in \cite{schuld-qml,schuld-sl}.
Also in the quantum state $|x\rangle$, the input signal $x_i$ can be any value.
So it contains a more general form than the model considered in \cite{schuld-qml}. 
Similarly, we can generate the quantum state $|w\rangle$ of the weight vector $\w=(w_1,\ldots,w_n)$.
The output of MLP can be any nonlinear function of $\x\cdot\w:=\sum_ix_iw_i$. However, computing $\x\cdot\w$ (or $\langle x|w\rangle$ when we only interested in the sign) is not difficult in a quantum computer due to swap test \cite{buhrman}. Moreover, any function of this value can be obtained by adding an ancilla qubit, just like HHL algorithm did \cite{harrow}. This shows another advantage of the representation $|x\rangle$ of the input signal vector $\x$.

However, reading the output is just the initial step of studying quantum MLP.
A big challenge of the quantum MLP is the learning algorithm because of the nonlinear structure of MLP.
Learning (or self-learning) ability is one of the most significant features of MLP. Many applications of MLP can be achieved by suitably constructing the network first.
Then let MLP to achieve the desired results from learning. So establish the learning algorithm of MLP in quantum computer is necessary and important.
In this paper, we will provide a method to integrate the nonlinear structure of MLP into quantum computer. At the same time, the good features of MLP (e.g., parallelism) 
and quantum computer (e.g., superposition) can be satisfied simultaneously. This provide us a new model to study MLP and its applications in quantum machine learning.

The integration is achieved by a new technique called parallel swap test proposed in a previous paper \cite{shao1} to study matrix multiplication. It is a generalization of swap test \cite{buhrman}, but in a parallel form. Simply speaking, parallel swap test can output the quantum information in parallel, which is required in the construction of quantum MLP. With this technique, we will show that the computation of the output signals and the learning algorithm (online and batch) of weights in quantum MLP can be achieved much faster (at least quadratic or even exponential faster) than the classical versions.
Finally, as an inspiration, we will study the learning (Hebb's learning rule) of Hopfield network.
This also achieves an exponential speedup at the number of input neurons over the classical learning algorithm.

The structure of this paper is as follows: In section \ref{Swap test in parallel}, we extend the swap test technique into a parallel form, which will play a central role in the simulation of quantum MLP. In section \ref{Preparation of quantum states}, we propose a technique to prepare quantum states, which is exponential fast than previous quantum algorithms. Although this cannot prepare the quantum states of all vectors efficiently, one can believe that it can solve many practical problems as fast as desired. In section \ref{Quantum simulation of Rosenblatt's perceptron}, we focus on the simulation of a perceptron. The simulation, which include the reading of output data and the back-propagation learning algorithm, of MLP in quantum computer will be studied in section \ref{Quantum simulation of multilayer perceptron}. Finally, in section \ref{Learning of Hopfield neural network in a quantum computer}, we study the learning algorithm of Hopfield neural network in a quantum computer.

{\bf Notation.} We will use bold letters $\x,\y,\z,\ldots$ to denote vectors and italic letters $|x\rangle,|y\rangle,|z\rangle,\ldots$ to denote their quantum states.
The norm $\|\cdot\|$ always refers to the 2-norm of vectors.

\section{Swap test in parallel}

\label{Swap test in parallel}

Swap test  was first proposed in \cite{buhrman} as an application of quantum phase estimation algorithm and Grover searching,
which can be used to estimate the probabilities or amplitudes of certain desired quantum states.
It plays an important role in many quantum machine learning algorithms to estimate the inner product of two quantum states.
In the following, we first briefly review the underlying problem swap test considers and the basic procedures to solve it.
Then, we review the generalized form of swap test considered in a previous work \cite{shao1} to study matrix multiplication in quantum computer.
This will play a central role in our construction of the model of multilayer perceptron in a quantum computer.

Let
\be\label{ini-form}
|\phi\rangle=\sin\theta|0\rangle|u\rangle+\cos\theta|1\rangle|v\rangle
\ee
be a unknown quantum state that can be prepared in time $O(T_{\rm in})$, where $|u\rangle,|v\rangle$ are normalized quantum states and $\theta$ is a unknown angle parameter.
The problem is how to estimate $\theta$ in quantum computer to precision $\epsilon$ with a high success probability.

Suppose that $|\phi\rangle$ comes from some other quantum algorithms, which means there is a given unitary $U$, which can be implemented in time $O(T_{\rm in})$, such that $|\phi\rangle=U|0\rangle$.
Let $Z$ be the 2-dimensional unitary transformation that maps $|0\rangle$ to $-|0\rangle$ and $|1\rangle$ to $|1\rangle$, which is usually called Pauli-Z matrix. Denote $G=(2|\phi\rangle\langle\phi|-I)(Z\otimes I)=U (2|0\rangle\langle0|-I)U^\dag(Z\otimes I)$, which is a rotation similar to the one used in Grover's algorithm. Then
\[
G=\left(
   \begin{array}{rr} \vspace{.2cm}
     \cos2\theta  &~~ \sin2\theta \\
     -\sin2\theta &~~ \cos2\theta \\
   \end{array}
 \right)
\]
under the basis $\{|0\rangle|u\rangle,|1\rangle|v\rangle\}$. The eigenvalues of $G$ are $e^{\pm\i2\theta}$ and the corresponding eigenvectors are
$|w_\pm\rangle=\frac{1}{\sqrt{2}}(|0\rangle|u\rangle\pm\i|1\rangle|v\rangle)$. Note that
$
|\phi\rangle=-\frac{\i}{\sqrt{2}} (e^{\i\theta}|w_+\rangle-e^{-\i\theta}|w_-\rangle ).
$
So performing quantum phase estimation algorithm on $G$ with initial state $|0\rangle^n|\phi\rangle$ for some $n=O(\log1/\delta)$.
We will get an good approximate of the following state
\be\label{final-form}
-\frac{\i}{\sqrt{2}}\Big[e^{\i\theta}|y\rangle|w_+\rangle-e^{-\i\theta}|-y\rangle|w_-\rangle\Big],
\ee
where $y\in \mathbb{Z}_{2^n}$ satisfies $|\theta-y\pi/2^n|\leq \epsilon$. The time complexity of the above procedure is $O(T_{\rm in}/\epsilon)$.
Perform a measurement on (\ref{final-form}), we will get an $\epsilon$ approximate of $\theta$.

Now let $|x\rangle,|y\rangle$ be two real quantum states that can be prepared in time $O(T_{\rm in})$. The requirement of quantum states to be real can be removed.
Then the above method provides us an quantum algorithm to estimate $\langle x|y\rangle$ to accuracy $\epsilon$ in time $O(T_{\rm in}/\epsilon)$. Actually, we just need to consider the state
\be \label{inner product:state}
|\phi\rangle=\frac{1}{\sqrt{2}}(|+\rangle|x\rangle+|-\rangle|y\rangle)
=\frac{1}{2}(|0\rangle(|x\rangle+|y\rangle)+|1\rangle(|x\rangle+|y\rangle)).
\ee
The probability of $|0\rangle$ (resp. $|1\rangle$) is $(1+\langle x|y\rangle)/2$ (resp. $(1-\langle x|y\rangle)/2$). So we can set $\sin\theta=\sqrt{(1+\langle x|y\rangle)/2}$ and $\cos\theta=\sqrt{(1-\langle x|y\rangle)/2}$. The quantum state $|\phi\rangle$ can be rewritten in the form (\ref{ini-form}) with $|u\rangle,|v\rangle$ correspond to the normalization of $|x\rangle+|y\rangle,|x\rangle-|y\rangle$. Therefore, the inner product $\langle x|y\rangle$ can be evaluated in time $O(T_{\rm in}/\epsilon)$ to precision $\epsilon$.
Note that if $|x\rangle,|y\rangle$ are complex quantum states, then the probability of $|0\rangle$ (resp. $|1\rangle$) is
$(1+\textmd{Re}\langle x|y\rangle)/2$ (resp. $(1-\textmd{Re}\langle x|y\rangle)/2$). So we can only get the value $\textmd{Re}\langle x|y\rangle$ by the above method.
However, the image part of $\langle x|y\rangle$ can be computed by considering the inner product of $|x\rangle$ with $\i|y\rangle$.
Concluding this, we get the following result, which is known as swap test \cite{buhrman}

\bp\label{cor:inner product}
Let $|x\rangle,|y\rangle$ be two quantum states, which can be prepared in time $O(T_{\rm in})$,
then $\langle x|y\rangle$ can be estimated to precision $\epsilon$ in time $O(T_{\rm in}/\epsilon)$.
\ep

As one can see swap test only returns the result about the inner product. One problem is that if there are $N$ such inner products we want to estimate, then we should apply swap test at least $O(N)$ many times. This will inevitably increase the whole complexity. In the following, we will extend swap test into a parallel form that can help us estimate all the inner products in parallel.

Let $f(y)$ be some functions such that $f(y)=f(-y)$ (i.e., $f$ is an even function), then from (\ref{final-form}), we can get
\be\label{final-form-1}
|f(\theta)\rangle|\phi\rangle,
\ee
by adding a register to store $f(\theta)$ and undoing the quantum phase estimation. This is a quantum state that we want to further make use of the quantum information about $\theta$ instead of outputting.
Moreover, from (\ref{inner product:state}) and (\ref{final-form-1}), we actually can obtain the following quantum state
\[
\frac{1}{\sqrt{2}}\Big(|0\rangle|x\rangle+|1\rangle|y\rangle\Big)\Big|f(\langle x|y\rangle)\Big\rangle,
\]
for any function $f$ in that cosine function is even.
The case for complex quantum states can be solved similarly by considering $\langle x|y\rangle$ and $\langle x|\i|y\rangle$ in parallel.
Concluding the above analysis, we have
\bp \label{prop-swap test}
Let $|x\rangle,|y\rangle$ be two quantum states, which can be prepared in time $O(T_{{\rm in}})$. Let $f$ be any function. Then there is a quantum algorithm runs in time $O(T_{{\rm in}}/\epsilon)$ to achieve
\be \label{swap test:general procedure}
\frac{1}{\sqrt{2}}(|0\rangle|x\rangle+|1\rangle|y\rangle)\mapsto
\frac{1}{\sqrt{2}}(|0\rangle|x\rangle+|1\rangle|y\rangle)|f(s)\rangle,
\ee
where $|\langle x|y\rangle-s|\leq \epsilon$.
\ep

From proposition \ref{prop-swap test}, it is easy to get the following results

\bt[Parallel Swap Test] \label{Parallel swap test}
Given $2N$ quantum states $|u_0\rangle,|v_0\rangle,\ldots,|u_{N-1}\rangle,|v_{N-1}\rangle$ which can be prepared in time $O(T_{\rm in})$ and $N$ functions $f_0,\ldots,f_{N-1}$.
Then there is a quantum algorithm with runtime $O(T_{\rm in}/\epsilon)$ to get the following quantum state
\be\label{Parallel swap test:eq}
\frac{1}{\sqrt{N}}\sum_{j=0}^{N-1}|j\rangle |f_j(s_j) \rangle,
\ee
where $|s_j-\langle u_j|v_j\rangle| \leq \epsilon$.
\et

\bo
The result (\ref{Parallel swap test:eq}) can be obtained by operating (\ref{swap test:general procedure}) in parallel because of the control qubit.
More precisely, construct the quantum state $\frac{1}{\sqrt{N}}\sum_{j=0}^{N-1}|j\rangle$ first. Then view $|j\rangle$ as a control qubit to prepare
$|\phi_j\rangle:=\frac{1}{\sqrt{2}} (|u_j\rangle|0\rangle+|v_j\rangle|1\rangle)$. So we can efficiently get a quantum state in the form $\frac{1}{\sqrt{N}}\sum_{j=0}^{N-1}|j\rangle|\phi_j\rangle$.
For each $|\phi_j\rangle$, perform the transformation (\ref{swap test:general procedure}) to get the information of $f_j(\langle u_j|v_j\rangle)$. So we have
$\frac{1}{\sqrt{N}}\sum_{j=0}^{N-1}|j\rangle|\phi_j\rangle|f_j(\langle u_j|v_j\rangle)\rangle$.
As discussed above, this is achieved by performing quantum phase estimation on a unitary operator $G_j$, which depends on $|\phi_j\rangle$. So the quantum state
$\frac{1}{\sqrt{N}}\sum_{j=0}^{N-1}|j\rangle|\phi_j\rangle|f_j(\langle u_j|v_j\rangle)\rangle$ is obtained by performing the quantum phase estimation on
$\sum_j|j\rangle\langle j|\otimes G_j$. Finally, we will get the desired state (\ref{Parallel swap test:eq}) by undoing the preparation of $|\phi_j\rangle$.
\eo

\bc
For any given quantum state $\sum_j\alpha_j|j\rangle$, which can be prepared in time $O(T_{{\rm in}})$ and any function $f$, we can obtain $\sum_j\alpha_j|j\rangle|f(\tilde{\alpha}_j)\rangle$ in time $O(T_{\rm in}/\epsilon),$ where $|\alpha_j-\tilde{\alpha}_j|\leq \epsilon$.
\ec
\bo
Denote $|\phi\rangle=\sum_j\alpha_j|j\rangle$ as the given quantum state, then $\alpha_j=\langle j|\phi\rangle$.
The desired result can be obtained in a similar way as the proof of Theorem \ref{Parallel swap test}.
\eo

\section{Preparation of quantum states}

\label{Preparation of quantum states}

For further application in the quantum model of multilayer perceptron, we discuss one method in this section to achieve quantum state preparation.
Let $\x=(x_0,\ldots,x_{m-1})$ be any complex vector, then its quantum state $|x\rangle = \frac{1}{\|\x\|} \sum_{j=0}^{m-1} x_j|j\rangle$ can be prepared by the following simple procedure (the idea comes from Clader et al. \cite{clader}):

Step 1, prepare $\frac{1}{\sqrt{m}} \sum_{j=0}^{m-1} |j\rangle$.

Step 2, apply control operator to get the component of $\x$, that is to prepare
\be \label{quantum-state-preparation}
\frac{1}{\sqrt{m}} \sum_{j=0}^{m-1} |j\rangle \left[ tx_j|0\rangle + \sqrt{1-t^2|x_j|^2}|1\rangle\right],
\ee
where $t=1/\max_j|x_j|$. The first part contains the information of $|x\rangle$, and $|0\rangle$ serves as a distinguish qubit. The complexity to get $|x\rangle$ is
$O(\sqrt{n}\max_j|x_j|(\log m)/\|\x\|)=O((\log n)\max_j|x_j|/\min_j|x_j|)$ when all components of $\x$ are nonzero.
We can overcome the case when $\min_j|x_j|=0$ by only considering the nonzero components of $\x$.
However, before performing any measurement, the complexity to get (\ref{quantum-state-preparation}) is $O(\log m)$.
Sometimes the quantum state (\ref{quantum-state-preparation}) that contains the information of $|x\rangle$ and the norm $\|\x\|$ is enough to solve certain problems.

A more efficient quantum algorithm is based on the linear combinations of unitaries technique (LCU for short), which was first proposed in \cite{shao1} (also see \cite{shao2} for more details).
The LCU problem can be stated as: given $m$ complex numbers $\alpha_j$ and $m$ quantum states $|v_j\rangle$, which can be prepared efficiently in time $O(T_{\rm in})$, where $j=0,1,\ldots,m-1$. Then how to prepare the quantum state $|y\rangle$ proportional to $\y=\sum_{j=0}^{m-1} \alpha_j |v_j\rangle$? And what is the corresponding complexity?
The following LCU idea was used to simulate Hamiltonian \cite{berry} and solve linear systems \cite{childs}.

Set $\alpha_j=r_je^{\i\theta_j}$, where $r_j>0$ is the norm of $\alpha_j$. Denote $s=\sum_{j=0}^{m-1}r_j$. Define the unitary operator $S$ as $S|0\rangle=\frac{1}{\sqrt{s}}\sum_{j=0}^{m-1}\sqrt{r_j}|j\rangle$.
Then $|y\rangle$ can be obtained from the following procedure:
Prepare the initial state $\frac{1}{\sqrt{s}} \sum_{j=0}^{m-1} \sqrt{r_j}|j\rangle|0\rangle$ by $S$. Then conditionally to prepare $|v_j\rangle$ according to the qubit $|j\rangle$, so we get $\frac{1}{\sqrt{s}} \sum_{j=0}^{m-1} \sqrt{r_j}e^{\i\theta_j}|j\rangle|v_j\rangle$. Finally, apply $S^\dag$ on $|j\rangle$, which yields $\frac{1}{s} |0\rangle \sum_{j=0}^{m-1} \alpha_j|v_j\rangle +\textmd{orthogonal parts}$.
It is easy to see that the complexity to obtain $|y\rangle$ equals $O((T_{{\rm in}}+\log m)s/\|\y\|)$. A direct corollary of this LCU is

\bp\label{state-preparation1}
For any vector $\x=(x_0,\ldots,x_{m-1})$, its quantum state can be prepared in time $O(\kappa(\x)\log m)$, where $\kappa(\x)=\max_k|x_k|/\min_{k,x_k\neq 0}|x_k|$.
\ep

\bo
Assume that all entries of $\x$ are nonzero, otherwise it suffices to focus on the nonzero entries of $\x$.
Then to prepare $|x\rangle$, one just need to choose $|v_j\rangle=|j\rangle$. At this time $O(T_{\rm in})=O(1)$. So the complexity is
$O((T_{{\rm in}}+\log m)\sum_j|\alpha_j|/\|\y\|)=O(\kappa(\x)\log m)$ since $\|\y\|\geq m\min_j|x_j|$ and $s\leq m\max_j|x_j|$.
\eo

The above result is the same as the result given by Clader et al. \cite{clader}.
Actually, based on the LCU given above, the quantum state can be prepared more efficiently.

\bp \label{state-preparation2}
Let $\x=(x_0,\ldots,x_{m-1})$ be a given vector, then the quantum state of $\x$ can be prepared in time $O(\sqrt{\log\kappa(\x)}\log m)$,
where $\kappa(\x)=\max_k|x_k|/\min_{k,x_k\neq 0}|x_k|$.
\ep

\bo
For simplicity, we assume that $|x_0|=\min_{k,x_k\neq 0}|x_k|$. Find the minimal $q$ such that $\kappa(\x)\leq 2^q$, so $q\approx \log \kappa(\x)$. For any $1\leq j\leq q$, there are several entries of $\x$ such that their absolute values lie in the  interval $[2^{j-1}|x_0|,2^{j}|x_0|)$. Define $y_j$ as the $m$ dimensional vector by filling these entries into the corresponding positions as them in $\x$ and zero into other positions. Then $\x=\y_1+\cdots+\y_q$. For any $j$, we have $\kappa(\y_j)\leq 2$, so the quantum state $|y_j\rangle$ of vector $y_j$ can be prepared efficiently in time $O(\log m)$ by proposition \ref{state-preparation1}. We also have
$|x\rangle=\lambda_1|y_1\rangle+\cdots+\lambda_q|y_q\rangle$, where $\lambda_j=\|\y_j\|/\|\x\|$. From the LCU method given above, the complexity to achieve such a linear combination to get $|x\rangle$ equals
$O((\log m)\sum_{j=1}^q {\|\y_j\|}/{\|\x\|}) = O(\sqrt{q}\log m) = O(\sqrt{\log\kappa(\x)}\log m),$
where the first identity is because of the relation between 1-norm and 2-norm of vectors, more precisely, it is a result of
$\sum_{j=1}^q \|\y_j\| \leq \sqrt{q} \sqrt{\sum_{j=1}^q \|\y_j\|^2}=\sqrt{q}\|\x\|$.
\eo

Compared with LCU, parallel swap test achieves a similar task except that the coefficients $\alpha_j$ are not given directly.
One main idea about parallel swap test we will frequently use in the following simulation of multilayer perceptron in quantum computer is that:
given $2l$ quantum states $|v_j^{\pm}\rangle$, then we can compute the inner product $\langle v_j^+|v_j^-\rangle$ in parallel. More precisely, prepare the following quantum state first
\be\label{initial-state}
\frac{1}{\sqrt{l}} \sum_{j=0}^{l-1} |j\rangle \otimes \frac{1}{\sqrt{2}} (|v_j^+\rangle|0\rangle+|v_j^-\rangle|1\rangle).
\ee
For each $\frac{1}{\sqrt{2}} (|v_j^+\rangle|0\rangle+|v_j^-\rangle|1\rangle)$, there is a unitary operator $G_j$ whose eigenvalues contain the information about
$\langle v_j^+|v_j^-\rangle$. So apply quantum phase estimation on $\sum_j|j\rangle \langle j|\otimes G_j$ with the initial state (\ref{initial-state}), we can get
\[
\frac{1}{\sqrt{l}} \sum_{j=0}^{l-1} |j\rangle \otimes \frac{1}{\sqrt{2}} (|v_j^+\rangle|0\rangle+|v_j^-\rangle|1\rangle) |\langle v_j^+|v_j^-\rangle\rangle.
\]

With this quantum state, we can perform any other desired operations we want about $\langle v_j^+|v_j^-\rangle$, such as apply control operation to put $\langle v_j^+|v_j^-\rangle$ into the coefficients as HHL algorithm did
\[\hspace{-1.5cm}
\frac{1}{\sqrt{l}} \sum_{j=0}^{l-1} |j\rangle \otimes \frac{1}{\sqrt{2}} (|v_j^+\rangle|0\rangle+|v_j^-\rangle|1\rangle) |\langle v_j^+|v_j^-\rangle\rangle
\Big[\langle v_j^+|v_j^-\rangle|0\rangle+\sqrt{1-\langle v_j^+|v_j^-\rangle^2}|1\rangle  \Big].
\]
Undo the quantum phase estimation to remove $|\langle v_j^+|v_j^-\rangle\rangle$, then one will obtain
\be\label{initial-state1}
\frac{1}{\sqrt{l}} \sum_{j=0}^{l-1} \langle v_j^+|v_j^-\rangle |j\rangle |0\rangle +
\frac{1}{\sqrt{l}} \sum_{j=0}^{l-1} \sqrt{1-\langle v_j^+|v_j^-\rangle^2} |j\rangle |1\rangle.
\ee

The above simple idea can help us manage $\langle v_j^+|v_j^-\rangle$ in parallel. More complicate procedures can also be obtained similarly.
The complexity to get (\ref{initial-state1}) is $O((\log l)/\epsilon)$. If we apply swap test to compute all $\langle v_j^+|v_j^-\rangle$ separately, then apply LCU to get (\ref{initial-state1}), the complexity will be $O(l/\epsilon)$. So in this problem, parallel swap test performs much better.

\section{Quantum simulation of Rosenblatt's perceptron}

\label{Quantum simulation of Rosenblatt's perceptron}

\subsection{Rosenblatt's perceptron}

The perceptron was first invented by Rosenblatt at 1958 \cite{rosenblatt58}, which was the first algorithmically described neural network.
The perceptron is also the simplest form of a neural network used for the (supervised) classification of patterns that are promised to be linearly separable.
The structure of a perceptron is very simple. It consists of $m$ input neurons with adjustable (synaptic) weights, a bias and an output neuron (see figure \ref{perceptron} below).

\begin{figure}[H]
  \centering
  \includegraphics[width=10cm]{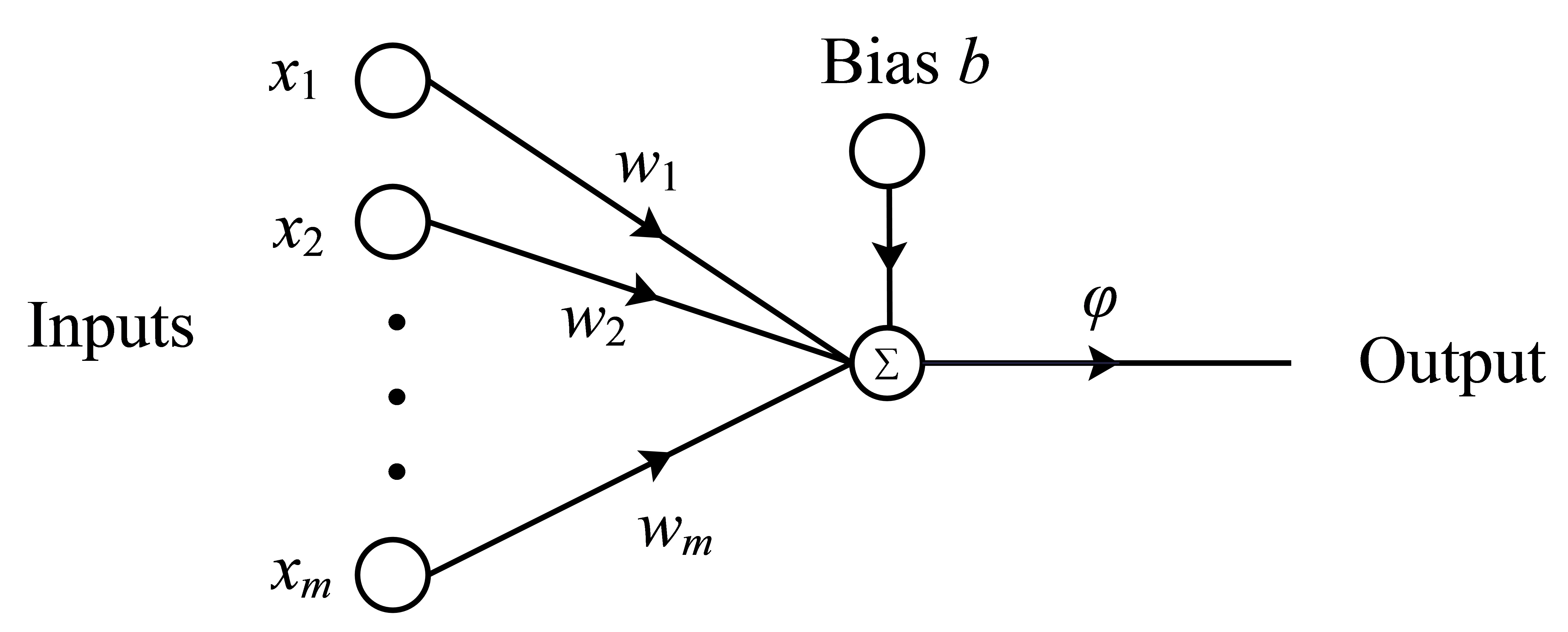}\\
  \caption{The graph of a perceptron: $x_1,\ldots,x_m$ are input signals, $w_1,\ldots,w_m$ are the weights, $b$ is the bias.
  The output is $\varphi(\sum_i x_iw_i+b)$, where $\varphi$ is a threshold function.}\label{perceptron}
\end{figure}

\begin{figure}[H]
  \centering
  \includegraphics[width=6cm]{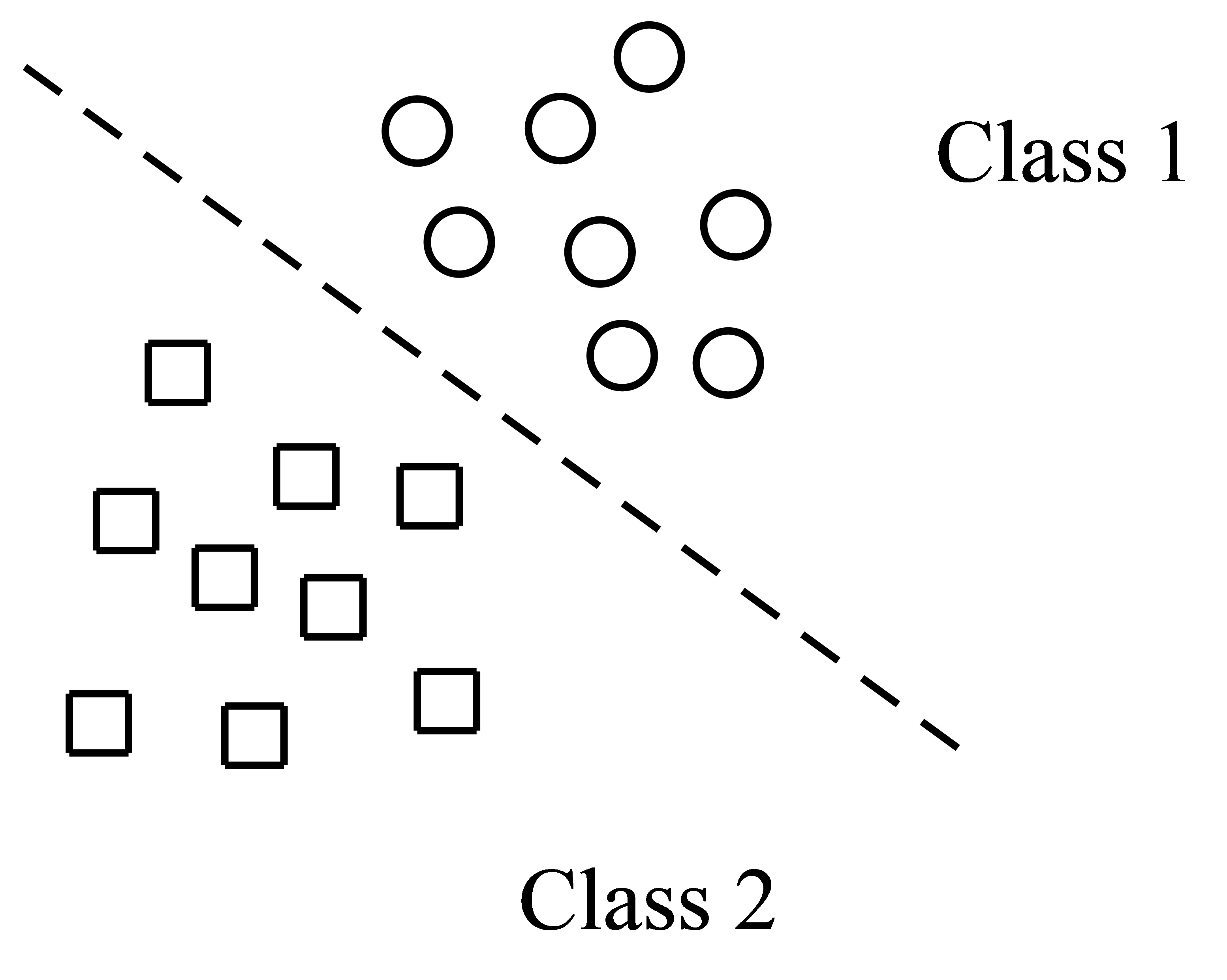}\\
  \caption{Linear separable training samples. There is a hyperplane to divide the two classes.}\label{Linear separable}
\end{figure}

Denote the input and the weight as $m+1$ dimensional vectors $\x=(x_0,x_1,\ldots,x_m)$ and $\w=(w_0,w_1,\ldots,w_m)$, where $x_0=1,w_0=b$.
The output is defined by a threshold function
\be \label{threshold function}
y=\varphi(\x\cdot\w)=\left\{
    \begin{array}{rl} \vspace{.2cm}
      1,  & \hbox{~~~~if $\w\cdot\x>0$;} \\
      -1, & \hbox{~~~~if $\w\cdot\x\leq0$.}
    \end{array}
  \right.
\ee
For classification problem, it means $y=1$ if $\x$ belongs to the first class; $y=-1$ if $\x$ belongs to the second class.
The hyperplane $\w\cdot\x=0$ defines the boundary of the classification (see figure \ref{Linear separable}).

Rosenblatt also defines a learning algorithm (or updating rule) to adjust the weight of the perceptron when it makes wrong decisions.
More precisely, given a set of training samples $\{(\x^t,r^t):t=0,1,\ldots,d-1\}$, where $r^t$ is the desired output of $\x^t$, that is
\be
r^t=\left\{
    \begin{array}{rl} \vspace{.2cm}
      1,  & \hbox{~~~~if $\x$ belongs to the first class;} \\
      -1, & \hbox{~~~~if $\x$ belongs to the second class.}
    \end{array}
  \right.
\ee
Then the updating rule is defined as
\be \label{updating rule}
\w^{\rm new} = \w^{\rm old} + \eta (r^t-y^t) \x^t
\ee
by randomly choosing a sample $(\x^t,r^t)$, where $\eta\in[0,1]$ is called the learning factor and $y^t=\varphi(\x\cdot\w^{\rm old})$.

Rosenblatt proved that this learning rule converges after a finite number iterations, which is currently known as perceptron convergence theorem.
Simply, from the updating rule (\ref{updating rule}), if $r^t=y^t$, then there is no updating. If $r^t>y^t$, that is $r^t=1,y^t=-1$, then
$\w^{\rm new} = \w^{\rm old} + 2\eta \x^t$. Since $r^t=1$, we know that $\x^t$ belongs to the first class.
However, $y^t=-1$ implies that we make the wrong decision to put $\x^t$ into the second class. So we should increase the value of $y^t$.
From the updated weight $\w^{\rm new}$, the new output of $\x^t$ equals $\varphi(\x^t\cdot \w^{\rm old} + 2\eta \x^t\cdot\x^t)$, which is larger than
the old output $\varphi(\x^t\cdot \w^{\rm old})$. Similar analysis also holds if $r^t<y^t$. More details about perceptron convergence theorem can be found in
\cite[Chapter 1]{haykin}. Such a learning algorithm is also called online learning \cite{alpaydin} in that the training only requires one sample at each step.
There are a lot of advantages of online learning, such as it saves the cost to store the training sample, it is easy to implement and so on
\cite[Chapter 11]{alpaydin}, \cite[Chapter 4]{haykin}.

\subsection{Simulation in quantum computer}

In this subsection, we study the corresponding quantum model of a perceptron, which include the reading of the output signal and the learning algorithm.
Due to the simple structure of a perceptron, the quantum model is also not complicate. In \cite{schuld-qml}, Schuld et al. proposed a quantum algorithm to
read the output signal by quantum phase estimation. Actually, get the output of a perceptron is not so difficult in a quantum computer, since it only needs to estimate
the inner product of two vectors. And swap test can achieve such a goal efficiently as described in section \ref{Swap test in parallel}.
In the following, we will mainly focus on the Rosenblatt's learning algorithm of a perceptron.

From the equation (\ref{threshold function}), the output $y$ only depends on the sign of $\w\cdot\x$. So we only need to focus on the normalized vectors of $\x$ and $\w$, that is the quantum states of $\x$ and $\w$. Define
\[
|x\rangle = \frac{1}{\|\x\|} \sum_{j=0}^m x_j |j\rangle, \hspace{.5cm}
|w\rangle = \frac{1}{\|\w\|} \sum_{j=0}^m w_j |j\rangle.
\]
These two quantum states can be prepared efficiently in quantum computer by the quantum algorithms given in section \ref{Preparation of quantum states}.
By swap test, the output $y=\varphi(\langle x|w\rangle)$ can be estimated to precision $\epsilon$ in time $O((\log m)/\epsilon)$.
For the updating rule (\ref{updating rule}), it can be achieved in the following way in a quantum computer:

\begin{description}
  \item[Algorithm 1: Training a perceptron in quantum computer]
  \item[Step 1] Randomly choose a sample $(\x^t,r^t)$.
  \item[Step 2] Compute $y^t=\varphi(\x\cdot\w^{\rm old})$ by swap test. If $r^t=y^t$ then go to {\bf Step 1}, else go to {\bf Step 3}.
  \item[Step 3] Prepare the quantum state $|w^{\rm new}\rangle$ of $\w^{\rm new}$ by the following procedure
  \begin{description}
    \item[Step 3.1] Prepare the quantum state $\frac{1}{\sqrt{2}} (|x^t\rangle|0\rangle+|w^{\rm old}\rangle|1\rangle)$.
    \item[Step 3.2] Apply control operator on it to prepare
          \[\ba{lll} \vspace{.2cm}
          \hspace{-1.8cm}& & \ds\frac{1}{\sqrt{2}}|x^t\rangle |0\rangle\Big[s\eta(r^t-y^t)\|\x\||0\rangle+\sqrt{1-s^2\eta^2(r^t-y^t)^2\|\x\|^2}|1\rangle\Big] \\
          \hspace{-1.8cm}&+& \ds\frac{1}{\sqrt{2}}|w^{\rm old}\rangle |1\rangle\Big[s\|\w^{\rm old}\||0\rangle+\sqrt{1-s^2\|\w^{\rm old}\|^2}|1\rangle\Big],
          \ea\]
          where $s=1/\max\{\eta|r^t-y^t|\|\x^t\|,\|\w^{\rm old}\|\}=1/\max\{2\eta\|\x\|,\|\w^{\rm old}\|\}$.
    \item[Step 3.3] Apply Hadamard transformation on the second register to get $\frac{s}{2}\|\w^{\rm new}\| |w^{\rm new}\rangle |0,0\rangle+{\rm orthogonal~part}$.
  \end{description}
  \item[Step 4] Go to {\bf Step 1} until converges.
\end{description}

In {\bf Step 3.3}, we do not need to perform measurements to get $|w^{\rm new}\rangle$ exactly. There are several advantages to perform no measurements in this step:

(1). Perform measurements will increase the complexity of the whole learning algorithm. Since the complexity to get $|w^{\rm new}\rangle$ is
$O(\max\{2\eta\|\x^t\|,\|\w^{\rm old}\|\}(\log m)/\|\w^{\rm new}\|)$, which can be very large. If we do not perform measurements, the complexity to get the
quantum state in {\bf Step 3.3} is just $O(\log m)$.

(2). The updating formula (\ref{updating rule}) also needs the information of $\|\w^{\rm old}\|$, which is already contained in the quantum state obtained in {\bf Step 3.3}.

(3). We can normalize $\x^t$ and $\w^{\rm old}$ in the initial step, so that $s=\max\{2\eta,1\}$ is small. This will cause no influences in the following iterations.
The reasons is that to compute $y^t$ based on the new weight, we can apply swap test to estimate the inner product between $|x^t\rangle$ and the quantum state in {\bf Step 3.3}.
The only influence is the factor $s/2$, which is a small constant lie between 1 and 2. It will not affect the output.

(4). Moreover, denote the quantum state in {\bf Step 3.3} as $|W^{\rm new}\rangle$. Then in the next step of iteration, we do not need to implement {\bf Step 3.1-3.3}, instead we can implement
\begin{description}
    \item[Step 3.1'] Prepare the quantum state $\frac{1}{\sqrt{1+s^2\eta^2}} (s\eta(r^t-y^t)/2|x^t\rangle|0,0\rangle|0\rangle+|W^{\rm new}\rangle|1\rangle)$.
    \item[Step 3.2'] Apply Hadamard operator on the last qubit, then we have
    \[\hspace{-1cm}\frac{s}{2\sqrt{1+s^2\eta^2}} \Big[\eta(r^t-y^t)|x^t\rangle+\|\w^{\rm new}\| |w^{\rm new}\rangle\Big]|0,0\rangle+{\rm orthogonal~part}.\]
\end{description}

As we can see, the coefficient is changed into $\lambda_2=\lambda_1/{\sqrt{1+4\lambda_1^2\eta^2}}$ from $\lambda_1=s/2$ after another step of iteration.
It is not hard to show that after $n$ steps of iteration, the coefficient becomes $\lambda_n=1/\sqrt{4s^{-2}+4(n-1)\eta^2}$. By the perceptron convergence theorem, assume that
the learning algorithm stops after $n$ steps of iteration, then the final quantum state has the form
$
\lambda_n \|\w^{\rm final}\| |w^{\rm final}\rangle |0\rangle+{\rm orthogonal~part}.
$
The complexity to get this state contains two parts:

(1). Prepare the quantum state in {\bf Step 3}: At the $j$-th step of iteration, the complexity is $O(j(\log m))$, since we perform no measurements.

(2). Estimating the value of $y^t$ by swap test: At the $j$-th step of iteration, the coefficient is $\lambda_j$. So to make the error of estimating $y^t$ is small in size $\epsilon$, the error chosen in swap test is $\epsilon/\lambda_j$. Then the complexity to estimate $y^t$ is $O(j^{3/2}\eta(\log m)/\epsilon)$ since the quantum state preparation costs $O(j(\log m))$ as discussed in (1).

Therefore, the final complexity of the learning algorithm of a perceptron in a quantum computer equals
\be \label{complexity0}
O(n^{3/2}(\log m)\eta/\epsilon+n(\log m)).
\ee
In a classical computer, estimating the inner product of $\x$ and $\w$ costs $O(m)$.
So the complexity of the learning algorithm of a perceptron after $n$ steps of iteration in a classical computer is $O(mn)$.
Compared to classical learning algorithm, the quantum algorithm achieves an exponential speedup at $m$.
As a compensation, its dependence on $n$ is worse than classical algorithm.

\section{Quantum simulation of multilayer perceptron}

\label{Quantum simulation of multilayer perceptron}

\subsection{Multilayer perceptron}

\begin{figure}[H]
  \centering
  \includegraphics[width=10cm]{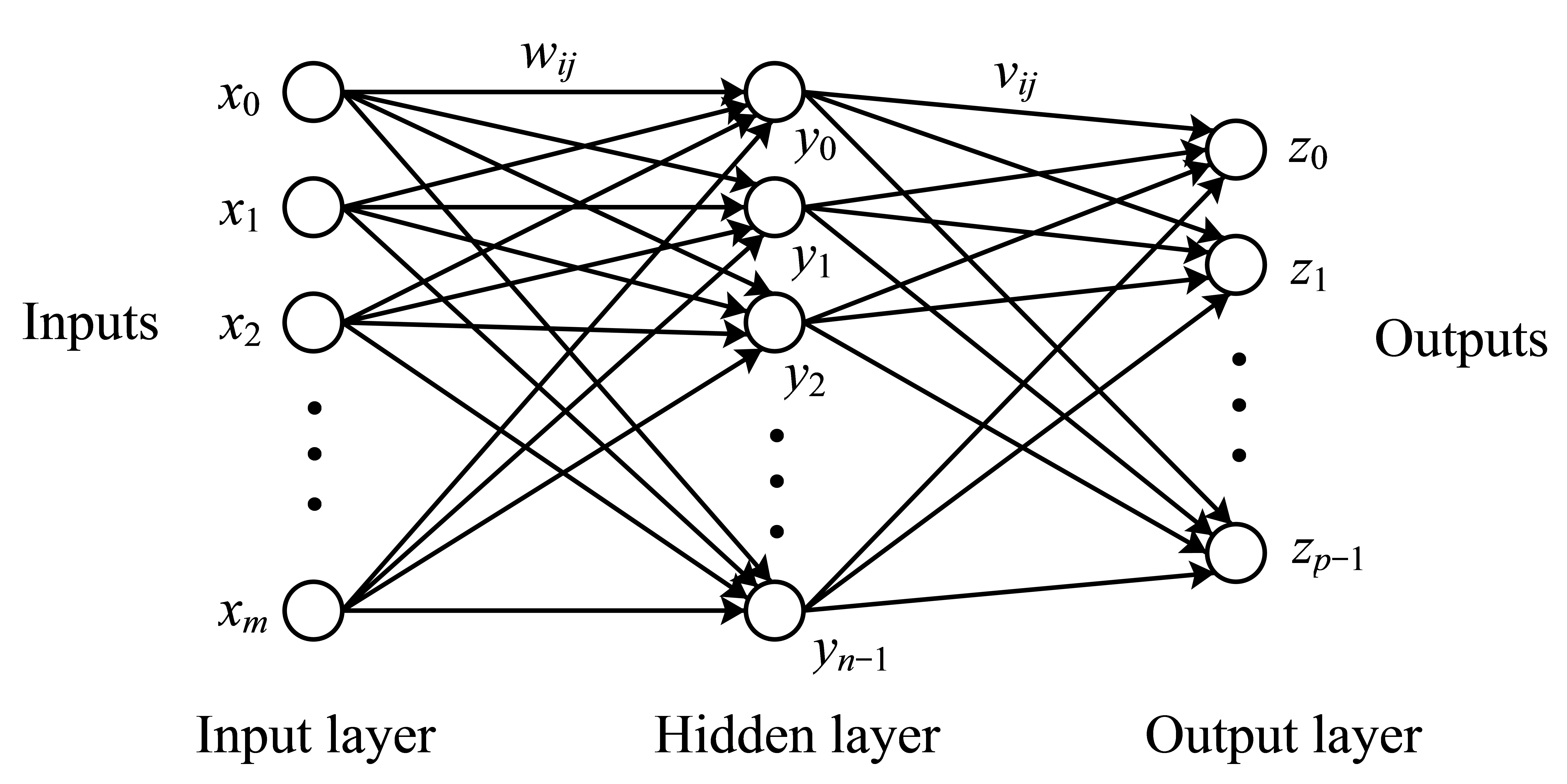}\\
  \caption{Graph of a multilayer perceptron with one hidden layer.}
  \label{multilayer-perceptron}
\end{figure}

Multilayer perceptron refers to a neural network with one or more hidden layers (see figure \ref{multilayer-perceptron}). It contains Rosenblatt's perceptron as a special case.
Perceptron can only used to perform the classification task with linear linearly separable patterns.
However, multilayer perceptron can accomplish much more complicate tasks, such as nonlinear classification, function approximation,
speech recognition, image recognition and so on \cite{haykin,jain}.
Instead of applying threshold function, multilayer perceptron often uses a nonlinear differentiable activation function.
Multilayer perceptron exhibits a high degree of interconnections, which makes its structure more complicate than perceptrons.
And this is one reason why it can achieve more complicate tasks than a perceptron.

The most commonly used activation function in multilayer perceptron is the sigmoid function defined as follows
\be\label{sigmoid function}
{\rm sigmoid}(a,\beta)=\frac{1}{1+\exp(-a\beta)}.
\ee
For any fixed $\beta$, the sigmoid function ${\rm sigmoid}(a,\beta)$ increases from 0 to 1 as $a$ grows from $-\infty$ to $+\infty$.
When $\beta$ goes into $+\infty$, the sigmoid function becomes closing to the threshold function.
In the following, we only focus on the activation function ${\rm sigmoid}(a):={\rm sigmoid}(a,1)$ with $\beta=1$.
%

If the input vector is $\x\in \mathbb{R}^m$, the weight vectors in the hidden layer are $\w_0,\ldots,\w_{n-1}\in\mathbb{R}^m$, then the $n$ outputs in the hidden layer are $y_i={\rm sigmoid}(\x\cdot\w_i)$, where $i=0,1,\ldots,n-1$. Denote $\y=(y_0,\ldots,y_{n-1})$.
Assume that the weight vectors in the hidden layer are $\v_0,\ldots,\v_{p-1}\in \mathbb{R}^n$, then the outputs are $z_j={\rm sigmoid}(\y\cdot\v_j)$, where $j=0,1,\ldots,p-1$.
Obviously, if $p<m$, then the multilayer perceptron performs a dimensionality reduction. It has been shown in \cite{bourlard} that a multilayer perceptron with one
hidden layer can implement principal components analysis, except that the hidden weights are not the eigenvectors sorted in importance but span the same space as the principal eigenvectors.

Training a multilayer perceptron is the same as training a perceptron; the only difference is that now each output is a nonlinear function of the inputs.
This makes the learning algorithm of a multilayer perceptron more complicate than a perceptron.
The learning algorithm in multilayer perceptron is called back-propagation, which was first discovered by Rumelhart, Hinton and Williams \cite{rumelhart} at 1986.
The development of the back-propagation algorithm represented a landmark in neural networks in that it provided a computationally efficient method for
the training of multilayer perceptrons. The reasoning of back-propagation learning algorithm is based on chain rule of calculus to compute the derivatives of composite functions.
Usually, the learning algorithm for the weights in the output layer is simple than the weights in the hidden layers.

Assume that there is only one hidden layer. Given a set of training samples $\{(\x^t,\r^t):t=0,1,\ldots,d-1\}$, where $\r^t\in \mathbb{R}^p$ refers to the desired output of $\x^t$ in the output layer. The error (called error energy averaged over the training sample, or the empirical risk) of this network is defined by $E = \frac{1}{2}\sum_{i,t} (\r_i^t-\z_i^t)^2$,
where $\z^t$ is the vector generated by $\z^t_i={\rm sigmoid}(\y^t\cdot\v_i)$ for all $i$ and $\y^t$ is the vector generated by $\y^t_j={\rm sigmoid}(\x^t\cdot\w_j)$  for all $j$.
The back-propagation learning algorithm is obtained by gradient descent method \cite{alpaydin, haykin} as follows:
In the gradient descent method, we need to compute the gradient of the error function with respect to $\w_i$ and $\v_j$. First, we see the computation of the gradient of $E$ with respect to $\v_j$. Direct calculation shows that
\[\ba{lllll}
\ds \frac{\partial E}{\partial \v_{jl}} &=& -\sum_{t} (\r_j^t-\z_j^t) \ds \frac{\partial \z_j^t}{\partial \v_{jl}}
&=&-\sum_{t} (\r_j^t-\z_j^t)  \z^t_j(1-\z^t_j) \y^t_l.
\ea\]
So the updating rule of $\v_j$ is
\be\ba{lll}\label{updating rule1}
\v^{\rm new}_j = \v^{\rm old}_j + \eta \sum_{t} \left[(\r_j^t-\z_j^t)  \z^t_j(1-\z^t_j)\right] \y^t.
\ea\ee
As for the gradient of $E$ with respect to $\w_i$, by chain rule, we have
\[\ba{lllll}
\hspace{-1.8cm} \ds \frac{\partial E}{\partial \w_{jk}} &=& -\sum_{i,t} (\r_i^t-\z_i^t) \ds \frac{\partial \z_i^t}{\partial \y^t_j} \frac{\partial \y^t_j}{\partial \w_{jk}}
&=&-\sum_{i,t} (\r_i^t-\z_i^t) \z^t_i(1-\z^t_i) \v_{ij} \y^t_j(1-\y^t_j) \x^t_k.
\ea\]
So the updating rule of $\w_j$ is
\be\ba{lll}\label{updating rule2}
\w^{\rm new}_j = \w^{\rm old}_j + \eta \sum_{t} \left[\sum_{i} (\r_i^t-\z_i^t) \z^t_i(1-\z^t_i) \v_{ij} \y^t_j(1-\y^t_j)\right] \x^t.
\ea\ee
Different choices of the activation function will lead to different updating rules of $\v_j$ and $\w_j$. In this paper, we only focus on the above two updating rules. All other choices can be studied similarly. The learning algorithm based on (\ref{updating rule1}) and (\ref{updating rule2}) is called batch learning, which uses all the sample at an epoch.
However, anther type of learning called online learning only uses on sample at one step.
In online learning, arbitrary choose a sample $(\x^t,r^t)$, then the error (called total instantaneous error energy) is defined by $E^t = \frac{1}{2}\sum_{i} (\r_i^t-\z_i^t)^2$.
Similarly analysis shows that the updating rule of online learning are
\be\ba{lll} \vspace{.2cm}\label{updating rule3}
\v^{\rm new}_j &=& \v^{\rm old}_j + \eta \left[(\r_j^t-\z_j^t)  \z^t_j(1-\z^t_j)\right] \y^t, \\
\w^{\rm new}_j &=& \w^{\rm old}_j + \eta \left[\sum_{i} (\r_i^t-\z_i^t) \z^t_i(1-\z^t_i) \v_{ij} \y^t_j(1-\y^t_j)\right] \x^t.
\ea\ee

\subsection{Simulation in quantum computer: reading the output data of each layer}

Different from a preceptron, the outputs of each layer and the learning algorithms of a multilayer perceptron depict strong nonlinear features. While quantum computer only allows unitary operations, which are linear operations. This is one reason why it is so hard to integrate the good features of multilayer perceptron and quantum computing to generate a useful model in quantum computer.
However, we will show in this and the next subsection that it is not so ``difficult" as it looks to integrate the nonlinear structure of multilayer perceptrons into a quantum computer.

In this subsection, we concentrate on the outputs reading of each layer.
First, we show how to get the output vector $\y$ of the first layer. Note that its $i$-th component equals $y_i={\rm sigmoid}(\x\cdot\w_i)$, where $i=0,1,\ldots,n-1$. If we compute all the components by swap test, then the complexity is at least $O(n)$. In the following, we give another method to construct the quantum data of $\y$. We will use $\varphi$ to denote the sigmoid function for simplicity in this subsection, although it has been used to denote the threshold function in Rosenblatt's perceptron. The procedure of preparing the quantum state of $\y$ can be stated as follows:

\begin{description}
\item[Algorithm 2:] {\bf Reading the output of the hidden layer in quantum computer}
\item[Step 1] Prepare $\frac{1}{\sqrt{n}} \sum_{i=0}^{n-1} |i\rangle.$
\item[Step 2] Apply control operator to prepare the following quantum state in parallel
\[
\frac{1}{\sqrt{n}} \sum_{i=0}^{n-1} |i\rangle \otimes \frac{1}{\sqrt{\|\x\|^2+\|\w_i\|^2}} \Big[\|\x\||x\rangle|0\rangle+\|\w_i\||w_i\rangle|1\rangle\Big].
\]
\item[Step 3] Apply parallel swap test on the above quantum state to generate
\be \label{state1}
\frac{1}{\sqrt{n}} \sum_{i=0}^{n-1} |i\rangle \otimes \frac{1}{\sqrt{\|\x\|^2+\|\w_i\|^2}} \Big[\|\x\||x\rangle|0\rangle+\|\w_i\||w_i\rangle|1\rangle\Big]
|\x\cdot \w_i\rangle.
\ee
\item[Step 4] Apply control rotation on the above quantum state and undo {\bf Step 2-3}
\be\ba{lll}  \label{state2} \vspace{.2cm}
& & \ds\frac{1}{\sqrt{n}} \sum_{i=0}^{n-1} |i\rangle  \Big[ \varphi(\x\cdot \w_i)|0\rangle+\sqrt{1-\varphi(\x\cdot \w_i)^2}|1\rangle \Big] \\
&=& \ds\frac{1}{\sqrt{n}} \sum_{i=0}^{n-1} \varphi(\x\cdot \w_i) |i\rangle |0\rangle+
\frac{1}{\sqrt{n}} \sum_{i=0}^{n-1} \sqrt{1-\varphi(\x\cdot \w_i)^2}|i\rangle|1\rangle.
\ea\ee
\end{description}

The explanation of the control operator in {\bf Step 2}: First prepare the quantum state
$|i\rangle \otimes \frac{1}{\sqrt{\|\x\|^2+\|\w_i\|^2}} [\|\x\||0\rangle+\|\w_i\||1\rangle]$.
Then conditionally preparing $|x\rangle$ and $|w_i\rangle$ based on the control qubit $|0\rangle$ and $|1\rangle$. Because of
the control register $|i\rangle$, these quantum states can be prepared in parallel in {\bf Step 2}.

In {\bf Step 3}, swap test returns a value $T_i$ such that
$|T_i-{\x\cdot \w_i}/{\sqrt{\|\x\|^2+\|\w_i\|^2}} | \leq \epsilon$, which lead to a good approximate of $\x\cdot \w_i$ with a small relative error.
The first part of (\ref{state2}) is the desired quantum state $|y\rangle$ of $\y$. Similar to the discussion in the quantum simulation of a perceptron in section \ref{Quantum simulation of Rosenblatt's perceptron}, we do not need to perform measurement to get $|y\rangle$ exactly. Therefore, the final complexity of the four steps is
\be \label{complexity1}
O((\log mn)/\epsilon).
\ee
This achieves an exponential speedup over the classical method, whose complexity is $O(mn)$.

\br
{\rm
In the following, any quantum state similar to (\ref{state2}) will be abbreviated written in the form
$\frac{1}{\sqrt{n}} \sum_{i=0}^{n-1} \varphi(\x\cdot \w_i) |i\rangle |0\rangle+|0\rangle^\bot$. That is we only show the part we are interested in. Here $|0\rangle^\bot$
contains two meanings: it means the hidden quantum state is orthogonal to the first part; it also means there exists a qubit that can help us distinguish the desired and undesired
quantum states. Sometimes, there may exist many qubits in $|0\rangle$ (such as $|0,0\rangle$), but we will still use $|0\rangle$ to denote it when the number of qubits is not important.
}
\er

The preparation of the quantum state of the vector $\z$ in the output layer is similar to above procedure. In this case, it will show the advantage of performing no measurements in (\ref{state2}). Denote the quantum state in \eref{state2} as $|Y\rangle$, then the preparation of the quantum state of $\z$ can be obtained in the following way:

\begin{description}
\item[Algorithm 3:] {\bf Reading the output of the output layer in quantum computer}
\item[Step 1] Prepare $\frac{1}{\sqrt{p}} \sum_{i=0}^{p-1} |i\rangle.$
\item[Step 2] Apply control operator to prepare
\[
\frac{1}{\sqrt{p}} \sum_{i=0}^{p-1} |i\rangle \otimes \frac{1}{\sqrt{1+\|\v_i\|^2}} \Big[|Y\rangle|0\rangle+\|\v_i\||v_i\rangle|0\rangle|1\rangle\Big].
\]
\item[Step 3] Apply parallel swap test on the above quantum state, which yields
\be \label{state3}
\frac{1}{\sqrt{p}} \sum_{i=0}^{p-1} |i\rangle \otimes \frac{1}{\sqrt{1+\|\v_i\|^2}}  \Big[|Y\rangle|0\rangle+\|\v_i\||v_i\rangle|0\rangle|1\rangle\Big]
|\y\cdot \v_i\rangle.
\ee
\item[Step 4] Apply control rotation on the above quantum state and undo {\bf Step 2-3}
\be \label{state4}
\frac{1}{\sqrt{p}} \sum_{i=0}^{p-1} \varphi(\y\cdot \v_i) |i\rangle |0\rangle+|0\rangle^\bot.
\ee
\end{description}

The whole procedure is the same as the preparation of the quantum information of $\y$. The only difference is {\bf Step 2}, now it becomes $|Y\rangle$ an $|v_i\rangle$.
Similarly, in {\bf Step 3}, we obtain a value $R_i$ such that
\be\label{relative-error2}
\left|R_i-\frac{1}{\sqrt{n}}\frac{\y\cdot \v_i}{\sqrt{1+\|\v_i\|^2}} \right| \leq \epsilon'
\ee
in time $O((\log mnp)/\epsilon\epsilon')$.
So to make $R_i\sqrt{n}$ an good approximate of $\y\cdot \v_i$ with small relative error $\epsilon$, we choose $\epsilon'$ such that
$\epsilon' \sqrt{n}=\epsilon$.
Therefore, the final complexity of the above procedure is
\be \label{complexity2}
O((\log mnp)\sqrt{n}/\epsilon^2).
\ee
Note that the classical algorithm to get the output has complexity $O(n(m+p))$.
So the quantum algorithm achieves an quadratic speedup at $n$ and exponential speedup at $m,p$. The quantum information of $\z$ lies in the first part of the quantum state (\ref{state4}).

\br\label{rmk}
{\rm
In the above construction, if we change the sigmoid function into the general form, then the complexity can be better.
More precisely, in the preparation of the quantum state of $\z$, if we change the function ${\rm sigmoid}(\y\cdot \v_i)$ into
${\rm sigmoid}(\x\cdot \w_i,1/\sqrt{n})$, then the complexity to prepare the quantum state (\ref{state4}) can be simply written as
$O((\log mnp)/\epsilon^2).$ This really achieves an exponential speedup over the classical method.
}\er

\subsection{Simulation in quantum computer: online learning algorithm}

Since online learning is simple than batch learning, so we first discuss the simulation of online learning (\ref{updating rule3}) in a quantum computer.
The updating rule (\ref{updating rule3}) for $\v_j$ needs the information of $\z^t_j=\varphi(\y^t\cdot\v_j)$ and $\y^t$.
Denote the quantum state in (\ref{state2}) by changing $\x$ into $\x^t$ as $|Y^t\rangle$, then the quantum information of $\y^t$ is
stored in $|Y^t\rangle$. The calculation of $\z^t_j$ depends on swap test, which is similar to (\ref{state3}).
Now we can describe the online learning algorithm of $\v_j$ as follows (the procedure is similar to the training of a perceptron):
\begin{description}
  \item[Algorithm 4:] {\bf Online training of the weights in the output layer in quantum computer}
  \item[Step 1] Randomly choose a sample $(\x^t,r^t)$.
  \item[Step 2] Prepare $|Y^t\rangle$ similar to (\ref{state2}).
  Then compute $\z^t_j=\varphi(\y^t\cdot\v_j^{\rm old})$ by applying swap test on
  $\frac{1}{\sqrt{1+\|\v_j^{\rm old}\|^2}}[|Y\rangle|0\rangle+\|\v_j^{\rm old}\||v_j^{\rm old}\rangle|0\rangle|1\rangle]$.
  If $\r^t_j\approx \z^t_j$ or $\z^t_j\approx 0$ or 1, then go to {\bf Step 1}, else go to {\bf Step 3}.
  \item[Step 3] Prepare the quantum state $|v_j^{\rm new}\rangle$ of $\v_j^{\rm new}$ by the following procedure
  \begin{description}
    \item[Step 3.1] Prepare the quantum state $\frac{1}{\sqrt{2}} (|Y^t\rangle|0\rangle+|v_j^{\rm old}\rangle|1\rangle)$.
    \item[Step 3.2] Apply control operator on it to prepare
          \[\ba{lll} \vspace{.2cm}
          \hspace{-3cm}& & \ds\frac{1}{\sqrt{2}}|Y^t\rangle |0\rangle\Big[s\eta (\r_j^t-\z_j^t)  \z^t_j(1-\z^t_j)|0\rangle
          +\sqrt{1-s^2\eta^2 [(\r_j^t-\z_j^t)  \z^t_j(1-\z^t_j)]^2}|1\rangle\Big] \\
          \hspace{-3cm}&+& \ds\frac{1}{\sqrt{2}}|v_j^{\rm old}\rangle |1\rangle\left[\frac{s\|\v_j^{\rm old}\|}{\sqrt{n}}|0\rangle
          +\sqrt{1-\frac{s^2\|\v_j^{\rm old}\|^2}{n}}|1\rangle\right],
          \ea\]
          where $s=1/\max\{\eta |\r_j^t-\z_j^t|  \z^t_j(1-\z^t_j),\|\w_j^{\rm old}\|\}\leq1/\max\{\eta/2,\|\v_j^{\rm old}\|/\sqrt{n}\}$.
    \item[Step 3.3] Apply Hadamard transformation on the second register to get $\frac{s}{2\sqrt{n}}\|\v_j^{\rm new}\| |v_j^{\rm new}\rangle |0\rangle+|0\rangle^\bot$.
  \end{description}
  \item[Step 4] Go to {\bf Step 1} until converges.
\end{description}

The complexity of one step of iteration is the same as (\ref{complexity2}), except the factor $\log p$. Because of the appearance of ${s}/{2\sqrt{n}}$ in the result of {\bf Step 3.3}
and note that $s$ is small, the iteration of the next step will be much easier as follows: Denote the quantum state in {\bf Step 3.3} as $|V_j^{\rm new}\rangle$.
\begin{description}
    \item[Step 3.1'] Prepare the quantum state
    \[\hspace{-1cm}
    \frac{2}{\sqrt{4+[\eta(\r_j^t-\z_j^t)\z^t_j(1-\z^t_j)]^2}}
    \left[\frac{s\eta (\r_j^t-\z_j^t)  \z^t_j(1-\z^t_j)}{2}|Y^t\rangle|0\rangle+|V_j^{\rm new}\rangle|1\rangle\right],
    \]
    \item[Step 3.2'] Apply Hadamard transformation on the second register, then we have
    \[\hspace{-3.4cm}
    \frac{s/\sqrt{n}}{\sqrt{4+[s\eta(\r_j^t-\z_j^t)\z^t_j(1-\z^t_j)]^2}}
    \Big[\|\v^{\rm new}_j\||v^{\rm new}_j\rangle + \eta \left[(\r_j^t-\z_j^t)  \z^t_j(1-\z^t_j)\right] \|\y^t\| |y^t\rangle \Big ]|0\rangle+|0\rangle^\bot.
    \]
\end{description}

The main costs of the above learning algorithm of $\v_j$ relies in the computation of $\z^t_j$, so after $N$ steps of iteration, the complexity is
\be \label{online-learning-complexity1}
O(N^{3/2}(\log mn)\sqrt{n}/\epsilon^2).
\ee
The reason of $N^{3/2}$ is the same as (\ref{complexity0}). The classical online learning algorithm of $\v_j$ has complexity $O(Nmn)$. Compared to the classical algorithm,
the quantum online learning algorithm achieves an quadratic speedup at $n$ and exponential speedup at $m$, however, the dependence on the number of iterations is worse.
If we apply the sigmoid function given in remark \ref{rmk}, then the corresponding quantum online learning algorithm can also achieve exponential speedup at $n$ over the classical online learning algorithm.

The online learning rule (\ref{updating rule3}) about $\w^{\rm new}_j$ needs the information of the summation $\sum_{i} (\r_i^t-\z_i^t) \z^t_i(1-\z^t_i) \v_{ij}$.
At this time, we need to calculate $\z^t_i$ in parallel. From (\ref{state3}), where we change $|Y\rangle$ into $|Y^t\rangle$, we can get
\be\hspace{-2cm} \label{online-learning-state}
\frac{1}{\sqrt{p}} \sum_{i=0}^{p-1} |i\rangle \otimes \left[
\sqrt{\tilde{s}(\r_i^t-\z_i^t) \z^t_i(1-\z^t_i) \v_{ij}}|0\rangle+\sqrt{1-\tilde{s}(\r_i^t-\z_i^t) \z^t_i(1-\z^t_i) \v_{ij}}|1\rangle
\right],
\ee
where $\tilde{s}=1/\max_i \{(\r_i^t-\z_i^t) \z^t_i(1-\z^t_i) \v_{ij}\}$. In the above, we assume that $(\r_i^t-\z_i^t) \z^t_i(1-\z^t_i) \v_{ij}>0$, otherwise we can add another qubit to deal with the negative parts. In the first part, the probability of $|0\rangle$ equals $\tilde{s}p^{-1} \sum_{i} (\r_i^t-\z_i^t) \z^t_i(1-\z^t_i) \v_{ij}$.
By swap test, this summation $\sum_{i} (\r_i^t-\z_i^t) \z^t_i(1-\z^t_i) \v_{ij}$ can be estimated in time
\be
O((\log mnp)\sqrt{np}/\epsilon^2)
\ee
to precision $\epsilon$. Similar to the learning algorithm of $\v_j$, except now $|Y^t\rangle$ is changed into $|x^t\rangle$, the online learning algorithm of $\w_j$ can
be implemented in quantum computer in time
\be
O(N^{3/2}(\log mnp)\sqrt{np}/\epsilon^2)
\ee
where $N$ is the number of iterations. The classical online learning algorithm about $\w_j$ has complexity $O(n(m+p))$. Therefore, the corresponding quantum learning algorithm achieves an exponential at $m$ and quadratic speedup at $n$ and $p$. The exponential speedup at $n,p$ can be achieved by choosing suitable sigmoid function as discussed in remark \ref{rmk}.

\subsection{Simulation in quantum computer: batch learning algorithm}

In this subsection, we focus on the modelling of batch learning algorithm in quantum computer. The whole idea is the same as the data reading of $\y$ and $\z$.
Due to the complicate expression of (\ref{updating rule1}), (\ref{updating rule2}), the corresponding quantum procedures are complicate too.
However, the procedure is very easy to understand.

First, we see how to achieve the updating rule (\ref{updating rule1}) of $\v_j$ in quantum computer. It contains a summation over $t=0,1,\ldots,d-1$, which can be achieved by considering (\ref{state2}) and (\ref{state4}) in parallel. The required quantum information in the updating rule include:
(1). Quantum state of $\y^t=(\varphi(\x^t\cdot\w_0),\ldots,\varphi(\x^t\cdot\w_{n-1}))$, which is contained in (\ref{state2}). The quantum state in (\ref{state2}) will be denoted as $|Y^t\rangle$.
(2). The inner product $\z^t_j=\varphi(\y^t\cdot\v_j)$, which can be obtained by applying swap test between $|Y^t\rangle$ and $|v_j\rangle$, see equation (\ref{state3}).
Now we can describe the basic idea of the updating rule of $\v_j$ in a quantum computer:

\begin{description}
\item[Algorithm 5:] {\bf Batch training of the weights in the output layer in quantum computer}
\item[Step 1] Prepare $\frac{1}{\sqrt{d}} \sum_{t=0}^{d-1} |t\rangle$.
\item[Step 2] Prepare $\y^t$ in parallel $\frac{1}{\sqrt{d}} \sum_{t=0}^{d-1} |t\rangle|Y^t\rangle$, where $|Y^t\rangle$ is the result obtained in (\ref{state2}) by changing $\x$ into $\x^t$.
\item[Step 3] Apply control operator to prepare
\[
\frac{1}{\sqrt{d}} \sum_{t=0}^{d-1} |t\rangle|Y^t\rangle\otimes \frac{1}{\sqrt{1+\|\v_j\|^2}} \Big[|Y^t\rangle|0\rangle+\|\v_j\||v_j\rangle|0\rangle|1\rangle\Big].
\]
\item[Step 4] Apply parallel swap test to obtain the information of $\z^t_j$, that is to prepare
\[
\frac{1}{\sqrt{d}} \sum_{t=0}^{d-1} |t\rangle|Y^t\rangle\otimes \frac{1}{\sqrt{1+\|\v_j\|^2}}
\Big[|Y^t\rangle|0\rangle+\|\v_j\||v_j\rangle|0\rangle|1\rangle\Big]|\y^t\cdot\v_j\rangle.
\]
\item[Step 5] Apply control rotation and undo the parallel swap test to prepare
\[
\hspace{-1.5cm}\frac{1}{\sqrt{d}} \sum_{t=0}^{d-1} |t\rangle|Y^t\rangle \Big[ s\eta (\r_j^t-\z_j^t)  \z^t_j(1-\z^t_j) |0\rangle
+ \sqrt{1-(s\eta (\r_j^t-\z_j^t)  \z^t_j(1-\z^t_j))^2} |1\rangle\Big]
\]
where $s=1/\max_t\eta (\r_j^t-\z_j^t)  \z^t_j(1-\z^t_j) \approx 1/4\eta$.
\item[Step 6] Apply Hadamard transformation on the first register $|t\rangle$
\[
\frac{s}{\sqrt{d^2n}} \sum_{t=0}^{d-1} \eta (\r_j^t-\z_j^t)  \z^t_j(1-\z^t_j) \|\y^t\| |y^t\rangle |0\rangle+|0\rangle^\bot.
\]
Denote this state of $|V_j\rangle$.
\item[Step 7] Prepare
\be\ba{lll} \label{state5}  \vspace{.2cm}
&&\ds\frac{1}{\sqrt{d^2n/s^2+\|\v^{\rm old}_j\|^2}}\left[\|\v^{\rm old}_j\||v^{\rm old}_j\rangle|0\rangle|
+\rangle+\frac{\sqrt{d^2n}}{s}|V_j\rangle|-\rangle\right]\\
&=& \ds \frac{1}{\sqrt{d^2n/s^2+\|\v^{\rm old}_j\|^2}} \|\v^{\rm new}_j\| |v^{\rm new}_j\rangle|0\rangle+|0\rangle^\bot.
\ea\ee
\end{description}

Similar to the complexity analysis of (\ref{complexity2}), the whole complexity to get the quantum state (\ref{state5}) is
\be \label{complexity4}
O((\log dmn)\sqrt{n}/\epsilon^2).
\ee

Next, we see the simulation of the updating rule (\ref{updating rule2}) of $\w_j$ in quantum computer. This requires more information than updating $\v_j$:
(1). All quantum states of $\x^t$, which can be prepared in advance.
(2). The summation $\sum_{i} (\r_i^t-\z_i^t) \z^t_i(1-\z^t_i) \v_{ij}$.
The underlying idea is similar, however, the description now is a little complicate than above because of the summation

\begin{description}
\item[Algorithm 6:] {\bf Batch training of the weights in the hidden layer in quantum computer}
\item[Step 1] Prepare $\frac{1}{\sqrt{d}} \sum_{t=0}^{d-1} |t\rangle \otimes \frac{1}{\sqrt{p}} \sum_{i=0}^{p-1} |i\rangle$.
\item[Step 2] Apply control operator to prepare
\[\ba{lll} \vspace{.2cm}
&& \ds\frac{1}{\sqrt{d}} \sum_{t=0}^{d-1} |t\rangle|x^t\rangle \otimes \frac{1}{\sqrt{\|\x^t\|^2+\|\w_i\|^2}} \Big[\|\x^t\||x^t\rangle|0\rangle+\|\w_j\||w_j\rangle|1\rangle\Big] \\
&& \ds \otimes \frac{1}{\sqrt{p}} \sum_{i=0}^{p-1} |i\rangle
\otimes \frac{1}{\sqrt{1+\|\v_i\|}} \Big[|Y^t\rangle|0\rangle+\|\v_i\||v_i\rangle|0\rangle|1\rangle\Big].
\ea\]
\item[Step 3] Apply parallel swap test and undo it to prepare
\be \label{update-rule-w-eq1}
\hspace{-2cm}\frac{1}{\sqrt{d}} \sum_{t=0}^{d-1} |t\rangle|x^t\rangle |\y^t_j\rangle \otimes \frac{1}{\sqrt{p}} \sum_{i=0}^{p-1} |i\rangle
\otimes \frac{1}{\sqrt{1+\|\v_i\|^2}} \Big[|Y^t\rangle|0\rangle+\|\v_i\||v_i\rangle|0\rangle|1\rangle\Big]|\z^t_i\rangle
\ee
\item[Step 4] Apply control rotation to prepare
\[\ba{lll}\vspace{.2cm}
&& \hspace{-2cm} \ds\frac{1}{\sqrt{d}} \sum_{t=0}^{d-1} |t\rangle|x^t\rangle |\y^t_j\rangle \otimes \frac{1}{\sqrt{p}} \sum_{i=0}^{p-1} |i\rangle
\Bigg[\sqrt{\tilde{s}(\r_i^t-\z_i^t) \z^t_i(1-\z^t_i) \v_{ij}}|0\rangle \\
&& \hspace{4cm} \ds+\sqrt{1-\tilde{s}(\r_i^t-\z_i^t) \z^t_i(1-\z^t_i) \v_{ij}}|1\rangle\Bigg],
\ea\]
where $\tilde{s}=1/\max_i (\r_i^t-\z_i^t) \z^t_i(1-\z^t_i) \v_{ij}\approx 1/4\max_i \v_{ij}$ and we assumed that $(\r_i^t-\z_i^t) \z^t_i(1-\z^t_i) \v_{ij}\geq0$,
otherwise we can give another copy to deal with the negative parts. The probability of $|0\rangle$ equals
$p^{-1}\tilde{s}\sum_{i=0}^{p-1} (\r_i^t-\z_i^t) \z^t_i(1-\z^t_i) \v_{ij}$.
\item[Step 5] Apply parallel swap test to prepare
\[\ba{lll}\vspace{.2cm}
&& \hspace{-2cm} \ds\frac{1}{\sqrt{d}} \sum_{t=0}^{d-1} |t\rangle|x^t\rangle |\y^t_j\rangle \otimes \frac{1}{\sqrt{p}} \sum_{i=0}^{p-1} |i\rangle
\Bigg[\sqrt{\tilde{s}(\r_i^t-\z_i^t) \z^t_i(1-\z^t_i) \v_{ij}}|0\rangle \\
&& \hspace{-.5cm} \ds+\sqrt{1-\tilde{s}(\r_i^t-\z_i^t) \z^t_i(1-\z^t_i) \v_{ij}}|1\rangle\Bigg] \otimes \Big|\sum_{i=0}^{p-1} (\r_i^t-\z_i^t) \z^t_i(1-\z^t_i) \v_{ij}\Big\rangle,
\ea\]
\item[Step 6] Undo {\bf Step 2-5} to prepare
\[\ba{lll} \vspace{.2cm}
&& \hspace{-3cm} \ds \frac{1}{\sqrt{d}} \sum_{t=0}^{d-1} |t\rangle|x^t\rangle
\otimes \Bigg[ \hat{s}\sum_{i=0}^{p-1} (\r_i^t-\z_i^t) \z^t_i(1-\z^t_i) \v_{ij}\y^t_j(1-\y^t_j) \|\x^t\| |0\rangle \\
&& \hspace{1cm} \ds+\sqrt{1-\Big(\hat{s}\sum_{i=0}^{p-1} (\r_i^t-\z_i^t) \z^t_i(1-\z^t_i) \v_{ij}\y^t_j(1-\y^t_j)\|\x^t\|\Big)^2} |1\rangle \Bigg],
\ea\]
where $\hat{s}=1/\max_t \{\sum_{i=0}^{p-1} (\r_i^t-\z_i^t) \z^t_i(1-\z^t_i) \v_{ij}\y^t_j(1-\y^t_j)\|\x^t\|\}\approx 1/4p\max_t\|\x^t\| \max_i \v_{ij}$.
\item[Step 7] Apply Hadamard transformation on the first register $|t\rangle$, which yields
\[
\frac{\hat{s}}{d} \sum_{t=0}^{d-1} \sum_{i=0}^{p-1} (\r_i^t-\z_i^t) \z^t_i(1-\z^t_i) \v_{ij}\y^t_j(1-\y^t_j)\|\x^t\| |x^t\rangle |0\rangle +|0\rangle^\bot
\]
Denote this state of $|W_j\rangle$.
\item[Step 8] Prepare
\be\ba{lll}\label{update-rule-w-eq5}  \vspace{.2cm}
&&\ds\frac{1}{\sqrt{d^2/\hat{s}^2+\|\w^{\rm old}_j\|^2}}\left[\|\w^{\rm old}_j\||w^{\rm old}_j\rangle|0\rangle|
+\rangle+\frac{d}{\hat{s}}|W_j\rangle|-\rangle\right]\\
&=& \ds \frac{1}{\sqrt{d^2/\hat{s}^2+\|\w^{\rm old}_j\|^2}} \|\w^{\rm new}_j\| |w^{\rm new}_j\rangle|0\rangle+|0\rangle^\bot.
\ea\ee
\end{description}

Next, we give a complexity analysis of the above procedures:
The first step costs $O(\log dp)$.
The costs in the second step is the preparation of $|Y^t\rangle$, which equals $O((\log mn)/\epsilon)$ by (\ref{complexity1}).
The third step needs to apply swap test to estimate $\z^t_i$. However, $|Y^t\rangle$ contains a coefficient $1/\sqrt{n}$ in (\ref{update-rule-w-eq1}). So the costs in this step is $O((\log dmnp)\sqrt{n}/\epsilon^2)$. The fourth step costs $O(1)$.
Because of the factor $1/\sqrt{p}$, the fifth step costs $O((\log dmnp)\sqrt{np}/\epsilon^2)$. The sixth step costs $O(1)$ and the seventh and eighth step cost $O(\log d)$. Therefore, the total cost of the eight steps is
\be
O((\log dmnp)\sqrt{np}/\epsilon^2).
\ee

Further updating of $\v_j$ and $\w_j$ are similar as above two algorithms. The difference is that in the following updating, we should apply
the weight vectors (\ref{state5}) and (\ref{update-rule-w-eq5}) instead of $|v_j\rangle,|w_j\rangle$ anymore. Assume that the learning algorithm stops at $N$
steps of iteration, then the complexities of updating $\v_j$ and $\w_j$ are
\be\label{complexity5}
O((\log dmn)\sqrt{n}/\epsilon^N)~{\rm and}~
O((\log dmnp)\sqrt{np}/\epsilon^N)
\ee
respectively. Similarly analysis as remark \ref{rmk}, the complexity (\ref{complexity5}) can be improved into
$O((\log dmn)/\epsilon^N)$ and $O((\log dmnp)/\epsilon^N)$ by choosing suitable sigmoid functions in each layer.
If the number $N$ of iterations is not large, then this achieves an exponential speedup over the classical birch learning algorithm,
whose complexity is $O(Ndmn)$ and $O(Ndn(m+p))$ respectively.

The complexity of batch learning algorithm (\ref{complexity5}) is exponential at the number $N$ of iterations.
Just like online learning, we can compute all the desired coefficients in advance. This can reduce the dependence of $N$ into polynomial,
however, the dependence on $d$, $m$, $n$, $p$ will be worse than (\ref{complexity5}).

\section{Learning algorithm of Hopfield neural network in a quantum computer}
\label{Learning of Hopfield neural network in a quantum computer}

Hopfield neural network (HNN) is one model of recurrent neural network \cite{hopfield82} with Hebb's learning rule \cite{hebb}. 
HNN serve as associative memory systems with binary threshold neurons. A variant form of HNN with continuous input range was proposed in \cite{hopfield84}.
The method introduced in this paper cannot directly applied to learn HNN in quantum computer. However, an amendatory version still exists. 
In this section, we will not focus too much on the details of learning algorithm of HNN in quantum computer. Also the reader can find more introductions about HNN in \cite[Chapter 13]{haykin,rojas}. A brief introduction about HNN can be found in \cite{schuld-qnn}. The idea about quantum HNN model applied here is different from \cite{rebentrost}, where they study HNN from density matrix and some related techniques (such as quantum principal component analysis and HHL algorithm).

Assume that $X=(x_{ij})_{P\times N}$ is a $P\times N$ matrices whose rows store the firing patterns of HNN. Denote the $i$-th row of $X$ as $\u_i$, the $j$-th column of $X$ as $\v_j$. The weight matrix $W=(w_{ij})_{N\times N}$ of HNN satisfies $w_{ij}=w_{ji}$ and $w_{ii}=0$. Denote the $j$-th column of $W$ as $\w_j$. The Hebb's learning rule of HNN is defined as
\be\label{Hebb learning rule}
w_{ij}=\frac{1}{P} \v_i\cdot\v_j.
\ee
And the update rule of $X$ is defined by
\be\label{HNN update rule}
x_{ij}=\varphi(\u_i\cdot\w_j),
\ee
where $\varphi$ is the threshold function (\ref{threshold function}) if all $x_{ij}$ take discrete values in $\{-1,1\}$ 
or the sigmoid function (\ref{sigmoid function}) if all $x_{ij}$ have continuous ranges.

Since the input of HNN is the matrix $X$, it is better to consider the following quantum state
\be\ba{lllll}
|X\rangle &=& \ds\frac{1}{\|X\|_F} \sum_{i=0}^{P-1}\sum_{j=0}^{N-1} x_{ij}|i,j\rangle &=& \ds\frac{1}{\|X\|_F} \sum_{i=0}^{P-1} \|\u_i\||i,u_i\rangle \\
&& &=& \ds \frac{1}{\|X\|_F} \sum_{j=0}^{N-1} \|\v_j\||v_j,j\rangle,
\ea\ee
where $\|X\|_F=\sqrt{\sum_{i,j}|x_{ij}|^2}$ is the Frobenius norm of $X$. Similarly, one can define $|W\rangle$.
We consider the problem to get the quantum state $|W\rangle$ by only giving $|X\rangle$.
Consider the quantum state
\be \label{HNN-eq0}
\frac{1}{N} \sum_{i,j=0}^{N-1} |i,j\rangle |+\rangle |X\rangle
=\frac{1}{N\|X\|_F} \sum_{i,j,k=0}^{N-1}\|\v_k\||i,j\rangle|+\rangle|v_k,k\rangle.
\ee
For any $i,j$, define
\[
\delta^{(0)}_{i,j}=\left\{
             \begin{array}{ll}
               1, & \hbox{if $i=j$;} \\
               0, & \hbox{if $i\neq j$.}
             \end{array}
           \right. \hspace{1cm}
\delta^{(1)}_{i,j}=\left\{
             \begin{array}{ll}
               1, & \hbox{if $i=j$;} \\
               2, & \hbox{if $i\neq j$.}
             \end{array}
           \right.
\]
As for $|i,j\rangle|0\rangle|v_k,k\rangle$, we change it into $|i,j\rangle|0\rangle|v_k,k\rangle|\delta^{(0)}_{i,k}\rangle$.
When $i=k$, we produce a copy of $|j\rangle$ to get $|i,j\rangle|0\rangle|v_i,i,j\rangle|1\rangle$.
Similarly, as for $|i,j\rangle|1\rangle|v_k,k\rangle$, we change it into $|i,j\rangle|1\rangle|v_k,k\rangle|\delta^{(1)}_{j,k}\rangle$.
And when $j=k$, we produce a copy of $|i\rangle$ to get $|i,j\rangle|1\rangle|v_j,i,j\rangle|1\rangle$. Perform these operations on (\ref{HNN-eq0}),
we obtain
\be\label{HNN-eq1}
\hspace{-2.5cm}
\frac{1}{\sqrt{2}N\|X\|_F} \sum_{i,j=0}^{N-1}|i,j\rangle\Big[|0\rangle (\|\v_i\||v_i,i,j\rangle|1\rangle+|\phi_0\rangle|0\rangle)
+|1\rangle (\|\v_j\||v_j,i,j\rangle|1\rangle+|\phi_2\rangle|2\rangle)\Big],
\ee
where $|\phi_0\rangle,|\phi_2\rangle$ are some undesired quantum states. Note that the inner product of $\|\v_i\||v_i,i,j\rangle|1\rangle+|\phi_0\rangle|0\rangle$ and
$\|\v_j\||v_j,i,j\rangle|1\rangle+|\phi_2\rangle|2\rangle$ equals $\|\v_i\|\|\v_j\|\langle v_i|v_j\rangle=P w_{ij}$. By proposition \ref{prop-swap test} and undo the above procedure, we will obtain
\be\label{HNN-eq2}
\hspace{-1cm}
\frac{P}{N\|X\|_F} \sum_{i,j=0}^{N-1}|i,j\rangle\Big[ sw_{ij}|0\rangle+\sqrt{1-s^2w_{ij}^2}|1\rangle \Big]
=\frac{Ps\|W\|_F}{N\|X\|_F} |W\rangle|0\rangle+|0\rangle^\bot.
\ee
where $s=1/\max_{i,j}|w_{ij}|$. The whole costs to get (\ref{HNN-eq2}) is $O((\log (PN))/\epsilon)$. The update rule (\ref{HNN update rule}) can be considered similarly
by considering $\frac{1}{\sqrt{2}}(|0\rangle|X\rangle+|1\rangle|W\rangle)$ instead of $|+\rangle|X\rangle$ in (\ref{HNN-eq0}).
When the number of iterations is not large, this algorithm achieves an exponential speedup over the classical Hebb's learning rule of HNN.
The above idea has been applied in \cite{shao1} to study matrix multiplication by parallel swap test, which achieves much better performance than SVE \cite{kerenidis} or HHL algorithm \cite{harrow}. Here, we also see that a similar idea also plays an important role in the learning of HNN in quantum computer.

\section{Conclusions}

In this paper, we establish a model of multilayer perceptron and a learning algorithm of Hopfield network in quantum computer based on parallel swap test technique.
The performance is much better than the classical algorithms when the number of layers or the number of iterations in not large.
On one hand, as for the parallel swap test technique, it play an important role in the design of the quantum model of multilayer perceptron and Hopfield network.
So as a research problem, it deserves to find more applications of parallel swap test.
Actually, we already found one such application in the Tikhonov regularization problem, which is used to deal with ill-posed inverse problem.
On the other hand, with this quantum model of multilayer perceptron and Hopfield network, it remains a problem to find its applications in quantum machine learning.
However, the learning algorithm in multilayer perceptron is based on gradient descent algorithm, which contains a low convergence rate. So it may deserve to consider
the learning algorithm based on Newton's method or quasi-Newton's method, where quantum computer can also achieve certain speedups under certain conditions.
Finally, it reserves as a problem to generalize this model to study the multilayer perceptron with more than two layers.

\ack
This work is supported by the NSFC Project 11671388 and the CAS Frontier Key Project QYZDJ-SSW-SYS022.

\end{document}